# Hierarchical short- and medium-range order structures in amorphous $Ge_xSe_{1-x}$ for selectors applications

Francesco Tavanti[1*], Behnood Dianat[1,2], Alessandra Catellani[1] and Arrigo Calzolari[1]

1 CNR-NANO Research Center S3, Via Campi 213/a, 41125 Modena, Italy
2 Dipartimento di Scienze e Metodi dell'Ingegneria, Università di Modena and Reggio Emilia, Via Giovanni Amendola, 2, 42122 Reggio Emilia, Italy
*corresponding author: francesco.tavanti@nano.cnr.it

**Abstract**

In the upcoming process to overcome the limitations of standard Von-Neumann architectures, synaptic electronics is gaining a primary role for the development of in-memory computing. In this field, Ge-based compounds have been proposed as switching materials for non-volatile memory devices and for selectors. By employing classical molecular dynamics, we study the structural features of both the liquid states at 1500K and the amorphous phase at 300K of Ge-rich and Se-rich chalcogenides binary $Ge_xSe_{1-x}$ systems in the range $0.4 \leq x \leq 0.6$. The simulations rely on a model of interatomic potentials where ions interact through steric repulsion, Coulomb and charge-dipole interactions given by the large electronic polarizability of Se ions. Our results indicate the formation of temperature-dependent hierarchical structures with short-range local orders and medium-range structures, which vary with the Ge content. Our work demonstrates that nanosecond-long simulations, not accessible via ab initio techniques, are required to obtain a realistic amorphous phase from the melt. Our classical molecular dynamics simulations are able to describe the profound structural differences between the melt and the glassy structures of GeSe chalcogenides. These results open to the understanding of the interplay between chemical composition, atomic structure and electrical properties in switching materials.

**Introduction**

A radical alternative to the existing scaling issues of Si-based technology relies on new disrupting solutions commonly referred as in-memory computing, where data are processed directly within the memory element integrated in the back-end-of-line (BEOL)[1]. Among the novel technologies which are being explored, the class of emerging memory devices or storage-class memories (SCMs), such as the resistive-switching memory (RRAM)[2], the phase-change memory (PCM)[3], and the spin-transfer torque memory (STTRAM)[4], are the most promising. This new class of memory technology has performance characteristics that fall between DRAM and flash characteristics, and bridges the latency gap between server storage and external storage. These devices have key characteristics, such as full compatibility with CMOS technology, simple geometry, low-power operations, and scalability beyond the 10nm node[5].
SCMs are in complex memory cells[6], which include 3D cross-point arrays of switching memories interconnected by selectors, which control the accessibility and quality of data transmission within and outside the entire cell[7]. Selectors are dynamical switches to the encoding mode with possible lowest bit error rate, which suppress the unselected shunt path for the leakage current in large (Gbit) arrays[8]. Selectors control the cell-to-cell write/read programming, and allow for the realization of high density crossbar memory array with $4F^2$ cell. The Ovonic Threshold Switching (OTS) selector technology showed the capability to overcome key issues with performance degradation over time, enabling high endurance and high ON/OFF current ratio ($I_{ON}/I_{OFF}$) required for crossbar applications[9,10]. The general OTS mechanism in amorphous chalcogenide materials is known[11] and consists in a volatile switch between a high resistive (OFF state) and a low resistive state (ON state) when the voltage applied to the material exceeds a critical voltage, called threshold voltage $V_{th}$. However, the detailed interplay between the parameters that characterize the materials (e.g. structure, composition, doping, chemical and thermal stability) and the influence they have on the device characteristics (e.g. data retention, power consumption and switching time) has not been completely understood.

Amorphous chalcogenides, such as $Ge_xSe_{1-x}$ compounds, have been proposed as ovonic switching materials[12] for both non-volatile memory devices and selectors[13], in view of their high ON-state drive current, good half-bias linearity, fast switching, endurance and a higher crystallization temperature respect to, e.g. GST or GeSbSeTe materials[14]. First experimental evidences indicate that the modulation of the stoichiometry ratio affects the electrical response of the device, in terms of both I-V characteristics and power dissipation[15]. In particular, devices based on Ge-rich GeSe compounds exhibit high OFF current, and low threshold voltage[16,17]; while Se-rich GeSe-based devices have higher energy gap, lower OFF-state leakage current and higher thermal stability[10]. The origin of this behavior is still unknown: while a lot of effort has been dedicated to GeTe as prototypical phase change material, or to $GeSe_2$ as efficient optoelectronic system, very little is known on quasi-stoichiometric $Ge_xSe_{1-x}$ compounds.

In analogy with other chalcogenide systems, it is expected that the characteristics of $Ge_xSe_{1-x}$ mobility gap and the presence of gap states is related to under/overcoordinated Ge atoms[14,15], and to the ratio of homopolar/heteropolar bonds in the sample[12]. On the experimental side, neutron or X-ray measurements only give statistical averages of the structures with the lack of description of the coordination path at the atomistic level. On the theoretical side, unraveling the interplay between chemical composition, local-order structures, trap states, and the resulting electrical response is a major challenge that requires the capability of treating large systems at the atomistic level along with their electronic structure. The prerequisite to this analysis is the identification of reliable atomic structures of the amorphous solid-state phase at room temperature. However, to date no univocal determination of structural properties of $Ge_xSe_{1-x}$ is available.

Most previous experimental and computational studies dealt with the $GeSe_2$ in both liquid and glass phases[18–22]. In the case of $Ge_xSe_{1-x}$, only studies on the liquid phase at high temperature (>700K) have been conducted, and no results are available for the amorphous solid phase at room temperature, especially for Ge-rich systems. The interpretation of the liquid $Ge_xSe_{1-x}$ is also non trivial, as no experiment or computer study was able to yield a definitive structure. A few quantum-mechanical models of amorphous GeSe and $GeSe_2$ systems exist[21,23,24], but are plagued by the severe computational costs that limit the size and the duration of the simulations. In this regard, the time required for reliable melting-and-quenching annealing cycles (>20ns) necessary to generate the amorphous glasses are usually not accessible to first-principles simulations.

In this paper, we investigate the structure of $Ge_xSe_{1-x}$ systems in the range $0.4 \leq x \leq 0.6$, by using classical Molecular Dynamics (MD) simulations. The use of classical MD raises up the possibility to study extended systems with thousands of atoms for several nanoseconds, which allows for medium-range spatial order organization. Our results indicate a clear effect of the stoichiometry in the formation of hierarchical short-range local-structures and medium-range connectivity networks in large scale amorphous models both in melt (high temperature) and glass (room temperature) phases. These are expected to play a crucial role in the definition of the mobility gap of the materials and their electrical response, thus explaining the differences observed experimentally in Ge-rich and Se-rich GeSe-based devices. The present analysis may also guide the identification of smaller but realistic models to be used in first-principles simulations, which are mandatory to correlate the complex structure of amorphous glasses with the corresponding electronic structure.

## Methods

*Computational details*

The $Ge_xSe_{1-x}$ systems have been built starting from the crystal structure of GeSe obtained from the Crystallography Open Database (COD), code: 4003515[25]. The $Ge_{50}Se_{50}$ initial structure has been built by replicating the unit cell 12, 4, 12 times along X, Y and Z direction respectively in order to obtain a system made by 4608 atoms with a density in the order of $5.5 \frac{g}{cm^3}$. The $Ge_{0.4}Se_{0.6}$ and the $Ge_{0.6}Se_{0.4}$ system have been built by substituting a proper number of atoms to obtain the correct stoichiometry. In order to simulate the correct amorphous structure, the density of the systems has been decreased by 8% with respect to the corresponding crystal phase as reported for the density of $GeSe_2$ and $Ge_xSe_{1-x}$[19,26]. In order to have a comparison for the density of the $Ge_{0.5}Se_{0.5}$, we performed an *ab initio* simulation at 1200K with variable cell dimensions coupled within the Parrinello-Rahman approach and with the Nosé-Hoover thermostat switched on electrons and ions obtaining a density in the order of $5.28 \frac{g}{cm^3}$, consistent with our thesis (see Supporting Information, SI, for further details).

The systems with the glass density have been melted at 1500K for 10ns in order to have a liquid phase and to let the systems lose the memory of the initial configuration. Then, the temperature has been gradually decreased to 300K with a cooling rate of 5K/ps, which is a good compromise to keep a quite low computational cost and to avoid freezing effects that can alter the medium-range structures[27]. After quenching, a production run has been conducted at 300K for 50ns. Simulations have been performed on an NVT ensemble with a Nosé-Hoover thermostat with a characteristic time of 10fs and the MD timestep was taken to be 1fs. Simulations have been performed using the Lammps package[28].

The Force Field (FF) employed is obtained customizing the well-known potential of Vashishta et al.[19,29,30] for $GeSe_2$ to the case of $Ge_xSe_{1-x}$. The implementation details of the generated FF are reported in the Supplementary Information (SI). The Vashishta FF has already been used to describe the structural features of several systems such as $Ag_2Se$[31], $Ag/Ge/Se$[32], $AlN$[33], $SiO_2$[34,35] and $GaAs$[36,37] with good accuracy; although ab initio molecular dynamics (AIMD) simulations provide higher accuracy, large systems and long simulation times as those required in this study are not accessible via AIMD.

*Structural analysis*

The Radial Distribution Function g(r) and the Faber-Ziman structure factor $S_n(q)$ have been computed on the last 5ns of the simulation at 1500K for the melt and on the last 15ns at 300K for the amorphous. The melting temperature for $Ge_{0.33}Se_{0.67}$ and $Ge_{0.5}Se_{0.5}$ is just above 1000K, while the glass transition temperature for $Ge_{0.4}Se_{0.6}$ at around 350K[30], thus the high and low temperature considered in the analysis are well representative of the melt and glassy states, respectively. The number of atoms bound to a central atom (i.e. the fold) has been computed using a cutoff distance of 3.0Å, with a tolerance of 0.2Å using an in-house code, called BELLO. For example, a 2-fold atom is an atom bound to the other two that lies within 3.0Å, while a 5-fold has 5 neighboring atoms at less than 3.0Å. Special cases are the 0-fold, which means that an atom is not bound to any other atom, and the 1-fold, where the atom is connected only to another atom, as for example at the end of a chain. The difference between tetrahedral and 4-fold atoms is given by the local order parameter, $q$[38,39]:

$$q = 1 - \frac{3}{8}\sum_{k>i}\left(\frac{1}{3} + \cos\theta_{ijk}\right)^2$$

where $\theta_{ijk}$ is the angle formed by atoms *i-j-k* with *j* the central atom. The local order parameter can assume values from 1, i.e. a perfect tetrahedral network, to 0 a six-fold octahedral network. We have chosen a range for the *q* value from 1 to 0.85 to describe the tetrahedral fold (high bond order) and from 0.85 to 0 for the generic 4-fold (low bond order), see **Figure 1**.

The rings statistics has been computed using the R.I.N.G.S. code[40] by using the King's criterion, which defines a ring as the shortest path connecting a number of atoms or nodes[41].

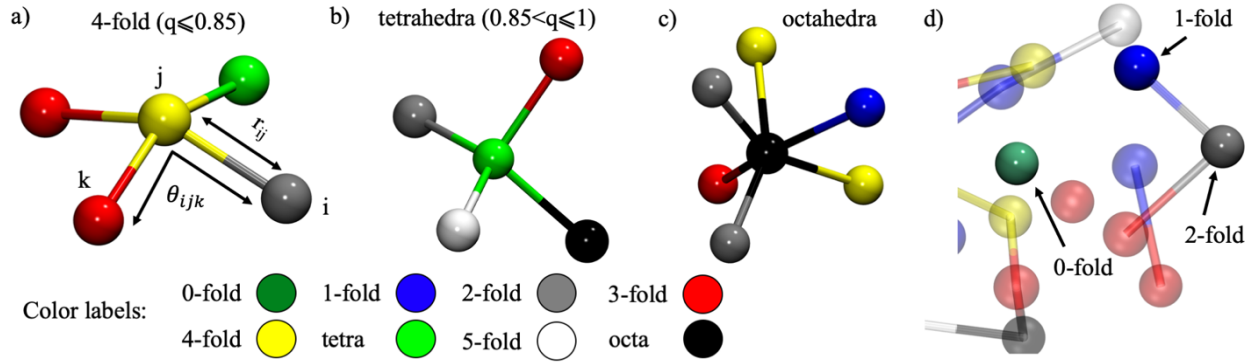

**Figure 1:** sample of configurations of the 4-fold, tetrahedral and octahedral structures, in panels a), b) and c), respectively. In panel d) the 0-fold, 1-fold and 2-fold atoms with respective connections scheme rendered as solid spheres, while the neighboring atoms as transparent ones. Below the color codes for each fold.

**Results and discussion**

*Short-range order*

The radial distribution function is obtained directly from MD simulations and it is used to analyze the two-body structural correlations. **Figure 2** shows the radial distribution function, g(r), for the amorphous at 300K of $Ge_xSe_{1-x}$. All previous experimental and theoretical studies dealt with either liquid and amorphous phases[18,20,42] in the case of $GeSe_2$ or liquid phase[22–24] in the case of $Ge_{0.5}Se_{0.5}$, while no results are available for the room-temperature amorphous structure of $Ge_xSe_{1-x}$ systems in the range of 0.4≤x≤0.6, that are used in selector devices. In this work, we used previous results of the g(r) of the liquid $Ge_{0.5}Se_{0.5}$ as a benchmark for the validation of our force field[43], see **Figure S1**.

The Se-rich system considered here ($Ge_{0.4}Se_{0.6}$) has composition close to the $GeSe_2$ (i.e. $Ge_{0.33}Se_{0.67}$). The resulting Ge-Se and Se-Se g(r)s of the melt phase have a behavior similar to previously published results[18,44] for $GeSe_2$ with primary peaks at 2.6Å and 4.0Å, respectively. A similar trend has been observed for both $Ge_{0.5}Se_{0.5}$ and $Ge_{0.6}Se_{0.4}$ liquid phases for Ge-Se and Se-Se bonds. The Ge-Ge g(r) for the liquid $Ge_{0.4}Se_{0.6}$ has a single main peak located at 4.2Å, without the presence of the first small peak at around 2.5Å that can be observed in *ab initio* simulation studies[24,45]. This difference can be ascribed to the different level of resolution of classical MD respect to *ab initio* MD and to the corresponding different timescales. Increasing the Ge content to 50%, the main peak splits into two with maximum located at 4.1Å and 5.35Å with a wider distribution as shown in **Figure S1**. Increasing the Ge content to 60%, we observed the formation of a new peak at 2.2Å, which is close to the Ge-Ge distance of 2.4Å in the crystal phase. This trend has been observed in particular for the liquid phase both using computational and experimental approaches[22,24,45] where the g(r) of the melt shows two peaks at 2.5 and 4Å, but at lower temperatures[22], see **Figure S1**. In particular, the work of Raty et al.[45] showed a quite different behavior of the g(r) of Ge-Ge going from 1053K to 1000K, while the work of van Roon et al.[24] showed a marked difference between the experimental and computational g(r) of Ge-Ge at the liquid phase. The high structural difference for the a-$Ge_xSe_{1-x}$ with x ≥ 0.5 respect to x=0.4 well agrees with the experimental FIR and Raman measurements by Trodahl[46]. We conclude that our FF well reproduces the structural features of the existing systems in the liquid phase.

In the amorphous phase at room temperature, the g(r) peaks appear more sharp and well-defined respect to the melt due to the lower mobility of atoms and to the increased structuring, as shown in **Figure 2**. For each stoichiometry, the position and the width of the Ge-Se and Se-Se peaks remain almost unaltered, while secondary peaks are found to be concentration-dependent. The case of Ge-Ge interactions is more complex. The 10% increased Ge content from the $Ge_{0.4}Se_{0.6}$ to the $Ge_{0.5}Se_{0.5}$ case results in the loss of a marked peak along with the formation of two peaks at 4.0 and 5.5Å, as seen for the melt phase, but with higher intensity. For $Ge_{0.6}Se_{0.4}$ the distribution is centered at around 3.76Å, with a marked secondary peak at 5.7Å. The shift of these distributions to smaller radii when increasing the Ge content is consistent with the values of pure Ge systems where the average Ge-Ge distance is 2.4Å[47]. Moreover, all three systems show a marked propensity for heteropolar Ge-Se bonds, while Ge-Ge and Se-Se are disfavored.

The integration over the first peak of the g(r) gives the coordination number of each chemical species contributing to the different peaks in the g(r) (see Figure S2). By averaging the elemental coordination number times their relative concentrations it is possible to obtain the Mean Coordination Number (MCN) of the system, which is relevant to empirically identify the rigidity of the chalcogenide glass: a floppy-to-rigid transition in chalcogenides is found for values of MCN of 2.4; values between 2.4 and 2.65 indicate that systems are characterized by flexible segments with a layer-like structure[48]. In the present case and assuming Ge atoms 4-coordinated and Se atoms 2-coordinated, the MCN ranges from 2.8 for $Ge_{0.4}Se_{0.6}$ to 3.2 for $Ge_{0.6}Se_{0.4}$ and indicates a 3D "stressed-rigid" phase[48]. The MCNs computed from our simulations are really close to the values predicted by the empirical law for $Ge_{0.4}Se_{0.6}$ and $Ge_{0.6}Se_{0.4}$ (2.80 and 3.04, respectively), while for $Ge_{0.5}Se_{0.5}$ the MCN computed from simulations is slightly higher (3.56) than the theoretical value of 3, as shown in Figure S2d.

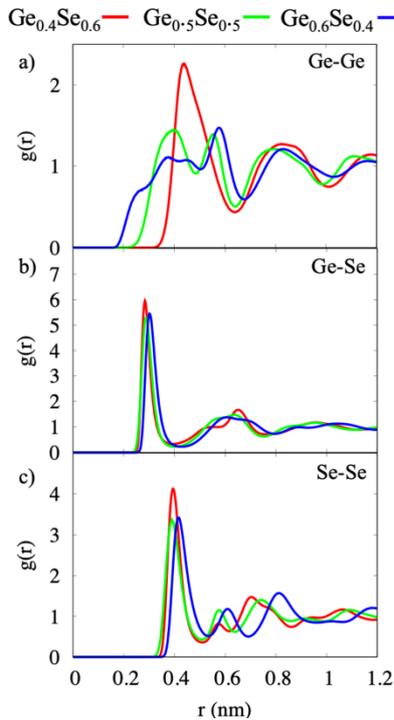

**Figure 2:** Radial distribution function, g(r), for the amorphous phase at 300K for each $Ge_xSe_{1-x}$ system. Panel a) shows the Ge-Ge, b) Ge-Se and c) Se-Se interactions.

**Figure 3** shows the results of the total neutron static structure factor $S_n(q)$ in the melt and amorphous phase. The First Sharp Diffraction Peak (FSDP) in the melt phase for the $Ge_{0.4}Se_{0.6}$ is located at 2.0Å$^{-1}$ with a second small peak at 3.4Å$^{-1}$ that are close to the values obtained for $GeSe_2$[29,49]. In the case of melt $Ge_{0.5}Se_{0.5}$, the FSDP is at 2.34Å$^{-1}$ with a small peak at 3.4Å$^{-1}$ and a minimum at 4.0Å$^{-1}$. These values are very close to those obtained both by spectrometer analysis and *ab initio* simulations and by Raty et al.[45] and Le Roux et al.[23] for liquid GeSe. In the amorphous phase, the FSDPs are much higher, due to the higher structuring of the systems, and the position of the first peaks is similar to those of the melt phase. The height of the peaks in the glassy state is slightly larger than in the molten phase, as previously shown for $GeSe_2$[19].

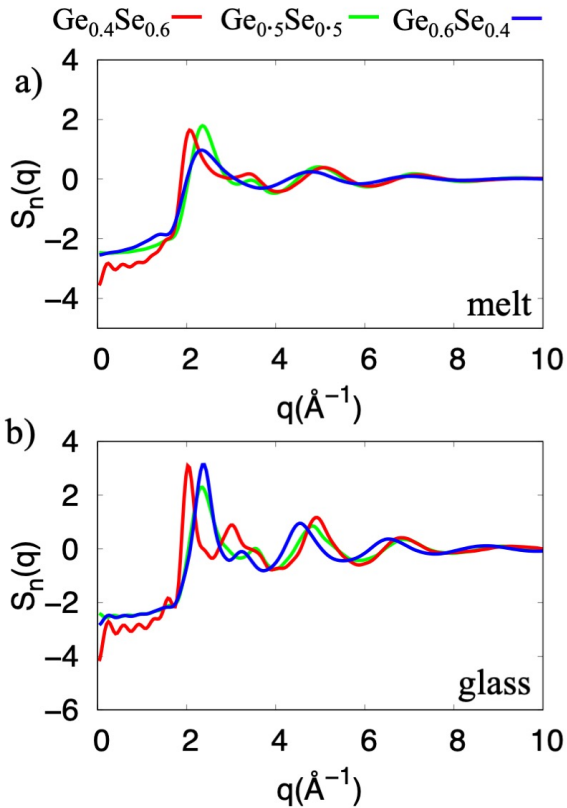

**Figure 3**: the Faber-Ziman total static structure factors for melt, panel a), and amorphous phases, panel b). Each composition is represented with a different color as shown in the legend above.

At high temperatures, the partial correlation function on $S_n(q)$ of Ge-Se has a minimum at 1.8Å$^{-1}$ together with a peak at 2.3Å$^{-1}$; while in the case of Se-Se, the FSDP is at 1.8Å$^{-1}$. These values are close to those obtained both by experiment and computer simulation studies on liquid GeSe[23,24,50]. Previous studies for the liquid $Ge_{0.5}Se_{0.5}$ system for Ge-Ge $S_n(q)$ showed a peak at around 1.7Å$^{-1}$, but the curve is more flat with respect to the others and also respect to *ab initio* simulations, whose results depend on the choice of the exchange-correlation functional[23,24,45,50]. The Ge-Ge interactions in the liquid reported in **Figure 4a** show a marked peak at 1.7Å$^{-1}$ with a trend similar to those previously obtained, confirming the good accuracy of the force field employed.

For all the amorphous phases, the FSDP has the same position as in the liquid phase, but the peaks are more pronounced, and the curve shapes change with stoichiometry. The most appreciable differences can be observed on $S_n(q)$ of Ge-Ge where the increase in the Ge content corresponds to decreasing FSDP height. In a similar way, the secondary peak shifts to smaller values of *q*, as shown in **Figure 4d**. From the intensities of the FSDP curves reported in **Figure 4**, it results that the main contribution to the total $S_n(q)$ is given by Ge-Ge interactions for amorphous $Ge_{0.5}Se_{0.5}$, while in the other systems Se-Se interaction is predominant.

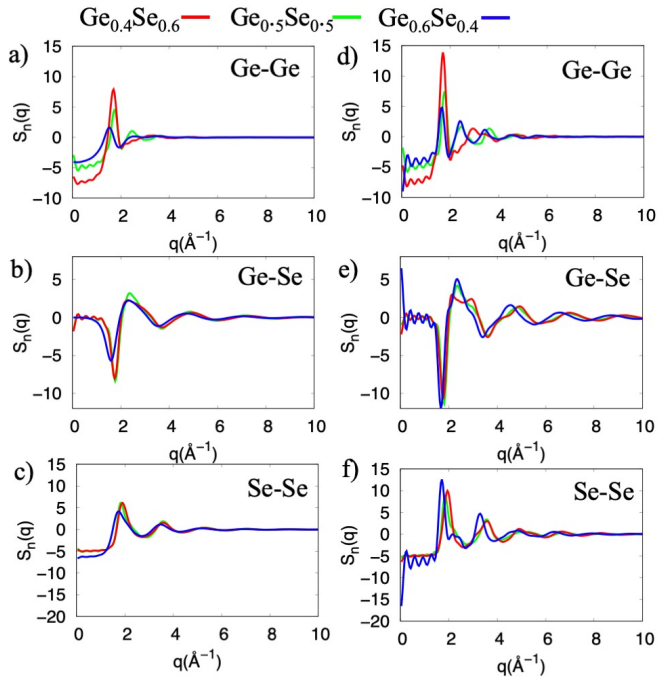

**Figure 4:** the Faber-Ziman partial static structure factors for melt, right panels, and amorphous phases, left panels. Each composition is represented with a different color as shown in the legend above. Panels a) and d) show the Ge-Ge $S_n(k)$, panels b) and e) the Ge-Se $S_n(k)$, panels c) and f) the Se-Se $S_n(k)$.

*Local order*

The understanding of local-order structures in terms of folding, composition and temperature evolution is of fundamental relevance in the design of materials for more efficient selectors. Local spatial-order in chalcogenides is expected to be closely related to their electronic properties[20,51,52]. Raty et al.[45] showed that mid-gap states are caused by Ge-Ge homopolar bonds of tetrahedral Ge, while Zipoli et al.[53] suggested that clustering of under-coordinated or over-coordinated Ge atoms are associated with mid-gap states. Moreover, it has been suggested that Ge-Ge chains in a crystalline-like environment produce mid-gap states[54] and that high-ordered structures, such as close but not-bound high-fold Ge atoms, are associated with the localization of mid-gap states[55].

**Figure 5** shows the percentage of folds during the simulation time, where the first 10ns are related to the melt phase and the last 50ns to the solid amorphous phase, assuming as central atom both Ge and Se. The local order in the liquid phase is preferentially given by low-folded structures with a high percentage of 1- and 2-folds and by the lack of high-folded structures characteristic of the glassy networks. After cooling down, the amorphous phase shows an increase in high-folded structures from 3- to 6-folds, while the bond distribution percentage depends on stoichiometry. In particular, the $Ge_{0.4}Se_{0.6}$ has a 2-, 3- and tetrahedral folds with a low amount of 5- and 6-folds with respect to the other two stoichiometries. On the contrary, the $Ge_{0.5}Se_{0.5}$ has the highest number of 4-, 5- and 6-folded structures, while the $Ge_{0.6}Se_{0.4}$ is in between, as shown in **Figure 5** and **Table 1**. These results clearly indicate that high Ge contents promote the formation of networks.

Experimental results based on photoemission XPS and UPS spectra on a-GeSe[56] suggest the presence of a chemically ordered structure and not a random bond structure. It has been shown that a thin film of a-GeSe deposited onto a cooled substrate has a 3:3 fold coordination (i.e. 3-fold Ge and 3-fold Se), but it relaxes onto a 4:2 after thermal annealing (i.e. 4-fold Ge and 2-fold Se). Other studies have also suggested a possible 4:2 folding of a-GeSe, but a common consensus is lacking. O'Reilly et al.[57] stated that a-GeSe has a 3:3 coordinated structure, while Trodahl et al.[58] used far infrared absorption to show that the 4:2 fold coordinated model is more appropriate for this system, although the 3:3 fold is also possible. Trodahl[46] also showed how heteropolar bonds are favored over homopolar bonds, as we observed in **Figure S3** and **S4**. We must highlight that these experimental studies do not have enough sensitivity to small density of defects because the signal is averaged over the whole material. Our findings for the a-$Ge_{0.5}Se_{0.5}$ reported in **Figure S3** and **S4** indicate that the most probable folds at room temperature are the 3- and 4-fold for both Ge and Se species.

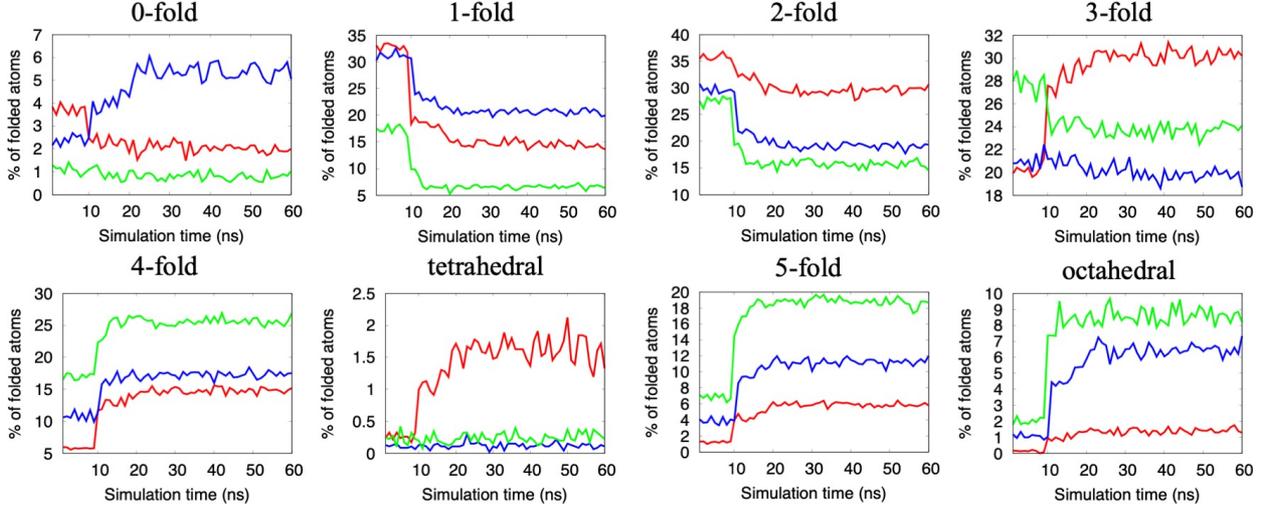

**Figure 5:** percentage of folded atoms for each $Ge_xSe_{1-x}$ system during the simulation time. The first 10ns correspond to the melt, the jump corresponds to the quenching, and the last 50ns to the amorphous phase. In red data for $Ge_{0.4}Se_{0.6}$, in green for $Ge_{0.5}Se_{0.5}$ and in blue for $Ge_{0.6}Se_{0.4}$.

In order to prove that the amorphous structure does not keep memory of the starting structure or of the melting-quenching preparation process, we performed a second cycle of heating at 1500K for 10ns and quenching to 300K by using as starting structure the last configuration reached from the amorphous phase (i.e continuing the MD simulation). The same parameters described in Methods were used and the atoms folding were calculated on the new obtained amorphous system. The resulting folding values for the liquid have the same average values reported in **Figure 5**, with a high number of low-folded atoms. This demonstrates that the system has completely lost the memory of the initial amorphous structure. In a similar way, after quenching, the a-GeSe show the same folding distributions as those reported in **Table 1**. Subsequent cycles of heating and quenching produce new amorphous structures with very similar properties but with no memory of previous configurations.
Apart the specific differences due to stoichiometry, the analysis of **Figure 5** indicates two general features: (i) the atomic folding increases as the temperature is reduced; (ii) the process to form highly folded structures requires a reorganization time of the order of ~10ns. The latter timescale is unaffordable for standard first principles simulations. Thus, fast ab initio quenching steps (~10-20 ps) may freeze the system in liquid-like phase (i.e. low-folded distribution), not allowing the atoms to rearrange and fold in the structures typical of the amorphous solid. As a comparison, we report in the SI the results of an ab initio Car-Parrinello molecular dynamics simulation on $Ge_{0.5}Se_{0.5}$ that clearly shows an overall low-folded atomic distribution of the sample, even at room temperature.

**Table 1:** percentage (%) of folded atoms for each $Ge_xSe_{1-x}$ amorphous system computed on the last 10ns of simulations. On the central cell the value considering both Ge and Se as central atoms, while we indicate the decomposition of the central Ge (Se) atom in the smaller cell at the top (bottom) right for each computed system.

|        | $Ge_{0.4}Se_{0.6}$ |     | $Ge_{0.5}Se_{0.5}$ |     | $Ge_{0.6}Se_{0.4}$ |     |
|--------|--------------------|-----|--------------------|-----|--------------------|-----|
| 0-fold | 1.9 ± 0.1          | 0.2 | 0.9 ± 0.2          | 0.3 | 5.3 ± 0.3          | 2.1 |
|        |                    | 1.7 |                    | 0.6 |                    | 3.2 |
| 1-fold | 14.2 ± 0.4         | 2.9 | 6.7 ± 0.3          | 3.2 | 20.6 ± 0.6         | 16.9 |

|              |              | 11.3 |              | 3.5  |              | 3.7  |
|              |              | 7.5  |              | 7.6  |              | 12.6 |
| 2-fold       | 29.9 ± 0.5   | 22.4 | 16.0 ± 0.4   | 8.4  | 19.0 ± 0.6   | 6.4  |
| 3-fold       | 30.4 ± 0.8   | 11.3 | 23.6 ± 0.5   | 11.1 | 19.8 ± 0.5   | 11.8 |
|              |              | 19.0 |              | 12.0 |              | 8.0  |
| 4-fold       | 14.7 ± 0.3   | 10.7 | 25.6 ± 0.7   | 12.4 | 17.3 ± 0.4   | 9.2  |
|              |              | 4.0  |              | 13.2 |              | 8.1  |
| Tetrahedral  | 1.5 ± 0.2    | 0.1  | 0.3 ± 0.1    | 0.2  | 0.1 ± 0.0    | 0.1  |
|              |              | 1.4  |              | 0.1  |              | 0.0  |
| 5-fold       | 5.9 ± 0.3    | 5.8  | 18.5 ± 0.4   | 9.7  | 11.3 ± 0.3   | 5.1  |
|              |              | 0.1  |              | 8.8  |              | 6.2  |
| Octahedral   | 1.5 ± 0.2    | 1.5  | 8.4 ± 0.4    | 5.5  | 6.5 ± 0.4    | 2.2  |
|              |              | 0.0  |              | 2.9  |              | 4.3  |

We analyzed the chemical distribution of the folded structures and we found that both Ge and Se may be at the center of the fold. However, while central Se atoms are connected only to peripheral Ge atoms (see **Figure S3** and **S4** in the Supporting Information for further details), central Ge atoms are connected both to Se and to Ge external atoms. This behavior is found for all three compositions, and in particular for the $Ge_{0.6}Se_{0.4}$. In this case, the central Ge atoms preferentially connect to Se, but always coordinate with at least one peripherical Ge per folded structure. For example, for the 4-fold case, Ge atoms mostly bind to 3 Se and 1 Ge, suggesting that Ge can form minority homopolar bonds, as previously shown by Liu et al.[59].

The analysis of the bond angle distribution (BAD) can help to gain insight into the short-range order structures, and in particular those for Ge-Se-Ge and Se-Ge-Se triads that are the most significant. As shown in **Figure 6**, the angle distributions are dependent on Ge concentration, but a general trend can be observed: the main peaks are located around 90°, with smaller contributions between 130° and 170°. Furthermore, increasing Se concentration, we detect an increase of spectral weight around 109.5° (**Table 1**), which is the typical fingerprint of tetrahedral structures, in agreement with a-$GeSe_2$ results[19,29]. On the contrary, increasing Ge content, the distribution is centered close to 90°, but the peak height diminishes and two additional peaks at 75° and at 130° appear. These findings suggest that increasing Ge content increase the disorder of the system in chalcogenides. The decomposition of BAD for each folded kind is reported in SI (**Figure S5**, **S6** and **S7**) and indicates that the major contribution to both angles in $Ge_{0.4}Se_{0.6}$ is associated to the 2- and 3-folds, while the contribution of tetrahedral and octahedral are negligible. In the case of $Ge_{0.5}Se_{0.5}$ the 3-, 4-, 5- and octahedral folds contribute almost equally to the distribution with marked peaks close to 90°. In the case shown by Trodhal[58] for the 4:2 fold coordination of a-GeSe, the $GeSe_4$ tetrahedral have an average angle of 90°, in agreement with **Figure 6b**.

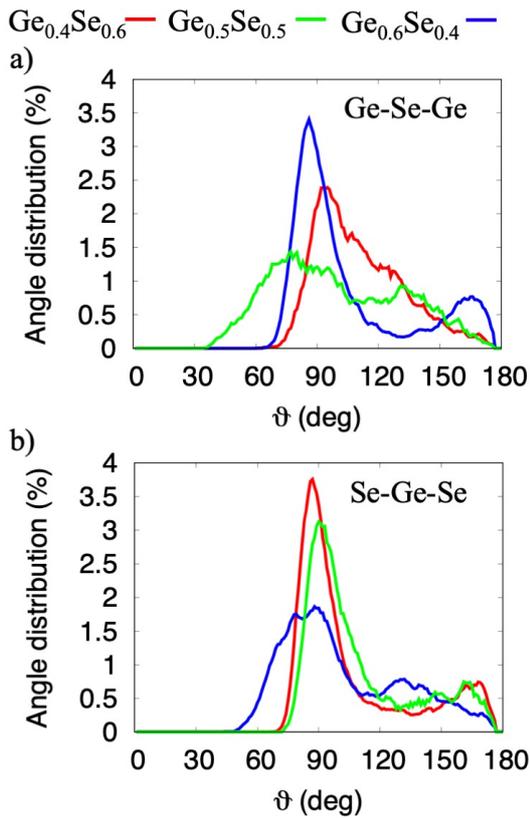

**Figure 6:** bond angle distribution of Ge-Se-Ge, panel a), and Se-Ge-Se, panel b), in amorphous $Ge_xSe_{1-x}$ systems at 300K.

*Medium-range order*

The medium-range order describing the connectivity of the network of $Ge_xSe_{1-x}$ has been obtained by computing the rings size and their spatial distribution using the King's algorithm[41], as implemented in the R.I.N.G.S. code[40]. Topological networks, such as amorphous systems, are represented in graph theory associating nodes for the atoms and links for the bonds. The possible formation of links between nodes is given by the analysis of the radial distribution function. A closed path connecting linked *n* nodes is called a *n-member* ring. Samples of rings can be found in **Figure S8**. Previous studies[60] on $GeSe_2$ showed a correlation between band gap and the presence of medium-range order structures, such as four- and six-membered rings and Ge-Se-Ge angles. These findings suggest that there is a strong correlation between the ring network and their electrical properties, although a clear view is still missing due to the high number of variables involved. The ring statistics have been performed using the location of the first peaks of the radial distribution functions as input for the ring's code. For each composition the locations are slightly different, and they implicitly take into account the different densities of the system. In this way, the results coming from the ring analysis are consistent and they do not depend on the chosen density.

The results of ring analysis are summarized in **Table 2**: Systems with a high Ge content ($Ge_{0.5}Se_{0.5}$ and $Ge_{0.6}Se_{0.4}$) may form both homopolar Ge-Ge and heteropolar Ge-Se bonds[61], that arrange in both even-membered and odd-membered rings. In particular, for $Ge_{0.5}Se_{0.5}$ the majority of the rings are composed of six, eight and ten elements (**Figure 7**), while the $Ge_{0.6}Se_{0.4}$ system has a predominance of three, and ten- to thirteen-member rings. In Se-rich $Ge_{0.4}Se_{0.6}$ system the lack of homopolar Ge-Ge and Se-Se bonds results in the formation of only even-membered rings (i.e. ABAB rings) with alternating sequences of Ge-Se atoms.

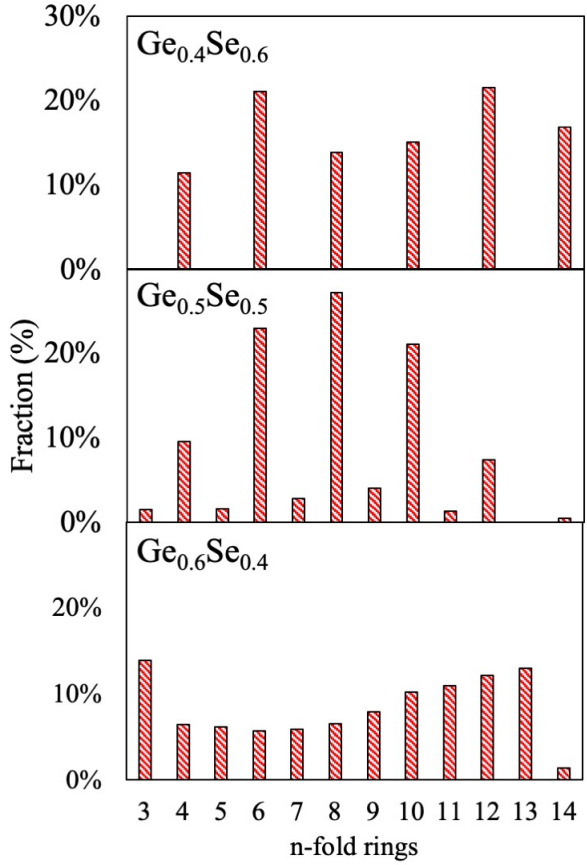

**Figure 7**: ring size probability for $Ge_xSe_{1-x}$ systems.

**Table 2:** ring size counts for $Ge_xSe_{1-x}$ systems.

| Ring size | $Ge_{0.4}Se_{0.6}$ | $Ge_{0.5}Se_{0.5}$ | $Ge_{0.6}Se_{0.4}$ |
|---|---|---|---|
| 3 | 0 | 63 | 402 |
| 4 | 46 | 403 | 184 |
| 5 | 0 | 69 | 178 |
| 6 | 85 | 965 | 165 |
| 7 | 0 | 118 | 169 |
| 8 | 56 | 1144 | 188 |
| 9 | 0 | 169 | 228 |
| 10 | 61 | 888 | 293 |
| 11 | 0 | 56 | 316 |
| 12 | 87 | 313 | 351 |
| 13 | 0 | 0 | 375 |
| 14 | 68 | 22 | 40 |
| Total | 403 | 4209 | 2889 |

The size of rings (i.e. number of nodes per ring) and their chemical character (i.e. homopolar or heteropolar links) are not sufficient to fully characterize the spatial distribution of rings across the sample and the possibility to have spatial networks of connected rings. In this regard, we introduce two indexes, namely Rc(n) and PN(n)[40]. Rc(n) is the number of Rings per Cell, that is the number of all different rings of size *n* that correspond at least once to the King's criterion[41], divided by the number of nodes; PN(n) is the Proportion of Nodes, which is the number of nodes that share more than one rings of size *n*. The former accounts for the spatial extension of rings, the latter for the connectivity of rings with the same size.

Rc(n) values (**Figure 8a**) for $Ge_{0.4}Se_{0.6}$ are extremely low for each ring size, indicating a limited propensity of the system in forming rings and a low network connectivity (low medium-range order). The other two systems have higher Rc(n) values, which correspond to a high probability to form rings, whose size distribution is given in **Table 2**. In $Ge_{0.5}Se_{0.5}$ the high Rc(n), along with the low PN(n) for n≤8 indicates the formation of connected networks made of rings with different sizes (i.e. 4-, 6- and 8-members rings). Notably, the PN(n) values reported in **Figure 8b** are lower than the ones estimated for other amorphous systems such as $GeS_2$, $SiO_2$[40], $TeO_2$[62] and $ZnCl_2$[63], indicating for these systems the formation of networks of interconnected rings with the same size (n-membered).

In order to have a deeper understanding of the network connectivity, we introduced other two parameters, $P_{min}(n)$ and $P_{max}(n)$, that describe the proportion of nodes for which the rings of size *n* are the shortest and the longest closed paths using these nodes to start the search, respectively. These parameters quantify the distribution of short-to-long connected rings. For homogenous networks made only of n-member rings, $P_{min}(n) = P_{max}(n) = 1$; while for networks that alternate short and long rings, $P_{min}(n)$ is higher for starting large-ring nodes, the opposite holds for $P_{max}(n)$.

In case of $Ge_{0.5}Se_{0.5}$ the $P_{min}$ curve (**Figure 8c**) decreases rapidly to 0 for n≥6, while $P_{max}$ raises to 1 for n=12 (**Figure 8d**). This corresponds to the formation of connected networks done by the alternation of rings of different sizes, ranging from n=6 to n=12. For both $Ge_{0.4}Se_{0.6}$ and $Ge_{0.6}Se_{0.4}$ a large $P_{min}$ value indicates the high probability of finding the shortest path for the node at the origin of the search, which corresponds to alternation of rings of all possible sizes. However, the different $P_{max}$ index indicates a different long-range distribution: low connectivity (low number of disconnected rings) for $Ge_{0.4}Se_{0.6}$ and high connectivity for $Ge_{0.6}Se_{0.4}$, in agreement with the corresponding high number of shared nodes (connected rings) expressed by Rc(n) and PN(n), respectively.

The large ring-network variability are expected to impart a different electrical behavior to $Ge_xSe_{1-x}$ systems, as a function of the stoichiometry, as detected in experimental measurements[10,16,17].

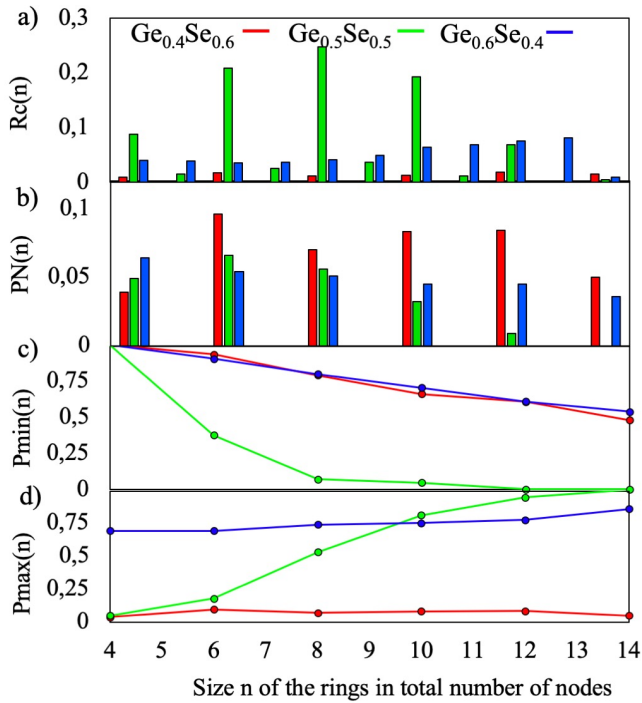

**Figure 8:** connectivity profiles for $Ge_xSe_{1-x}$ system: Rc(n) in a), PN(n) in b), $P_{min}(n)$ in c) and $P_{max}(n)$ in d).

**Conclusions**

In summary, we have performed classical molecular dynamics (MD) simulations of large-scale $Ge_xSe_{1-x}$ systems in the $0.4 \leq x \leq 0.6$ range to study the structure of their liquid and amorphous phases over ns-long timescales. MD simulations well reproduce all the main short-range features previously observed by *ab initio* molecular dynamics and by neutron diffraction studies for similar systems, such as $GeSe_2$ and liquid $Ge_{0.5}Se_{0.5}$, and provide a reliable model for the amorphous phase at room temperature. Amorphous phases are characterized by different levels of ordering depending on the relative Ge content. On the short-range, all systems present a predominant amount of highly folded structures, whose chemical composition differ in the formation of homopolar Ge-Ge bonds, for Ge-rich samples. On the medium-range, systems with Ge from 50% to 60% show the formation of a dense and interconnected network with a high percentage of rings, while networks of systems containing lower Ge content are characterized by a reduced number of disconnected rings.

In analogy with other chalcogenide systems, these structural differences are expected to affect the gap properties of $Ge_xSe_{1-x}$ compounds. However, while a different hierarchical short- and long-range order is clearly seen in three examined stoichiometries, its direct effect of the electronic and electrical properties (e.g. ovonic switching) has to be confirmed by specific quantum-mechanical investigations. The success of classical MD simulations in the description of amorphous $Ge_xSe_{1-x}$ structures paves the way to the study of other chalcogenide glasses, including dopants, for applications as OTS chalcogenides materials.

**Supporting Information**
SI includes the numerical details and tests on the generated FF (Section **S1**); further analysis of $Ge_xSe_{1-x}$ melt and amorphous phase, including g(r) for the liquid phase, coordination numbers atom statistics of the folds, decomposition of angle distribution, and sample of rings structures (Section **S2**); and the comparison with an ab initio molecular dynamics simulation (Section **S3**).

**Corresponding Author**
*To whom correspondence should be addressed
E-mail: francesco.tavanti@nano.cnr.it

**Author Contributions**
F.T. performed the simulations and data analysis. B.D., A.C., A.C. and F.T. planned the computational experiments and discussed results. The manuscript was written through contributions of all authors. All authors have given approval to the final version of the manuscript.

**Acknowledgments**
We thank S. Clima and P. D'Amico for fruitful discussions.

**Funding**
This work is funded by EC through H2020-NMBP-TO-IND project GA n. 814487 (INTERSECT).

# Supporting information

## Hierarchical short- and medium-range order structures in amorphous Ge$_x$Se$_{1-x}$ for selectors applications

**Francesco Tavanti[1*], Behnood Dianat[1,2], Alessandra Catellani[1] and Arrigo Calzolari[1]**

1 CNR-NANO Research Center S3, Via Campi 213/a, 41125 Modena, Italy
2 Dipartimento di Scienze e Metodi dell'Ingegneria, Università di Modena and Reggio Emilia, Via Giovanni Amendola, 2, 42122 Reggio Emilia, Italy
*corresponding author: francesco.tavanti@nano.cnr.it

## S1. Force Field

The force field (FF) adopted in the present work has been derived from the well-known potential of Vashishta et al.[1–3] for GeSe$_2$. The generic form of the FF is given by three terms:

$$\phi(r) = \frac{Z_i Z_j}{r} - \frac{\frac{1}{2}(\alpha_i Z_j^2 + \alpha_j Z_i^2) e^{-\frac{r}{r_{4s}}}}{r^4} + \frac{H_{ij}}{r^{\eta_{ij}}}.$$

The first term represents the Coulombic interaction where $Z_i$ is the charge of the $i$th particle. The charges give rise also to charge-dipole potential due to the high polarizability of large ions (Se), which results in the second term of the potential where $\sigma_i$ represent the polarizability of the $i$th particle and $r_{4s}$ is a decay length, i.e. 4.43Å, that screens the charge-dipole interaction. As in the original FF, the polarizability of Ge$^{4+}$ is 0 due to the smaller size respect to Se$^{2-}$, while the polarizability of Se$^{2-}$ is 7Å$^4$. The last term describes a steric repulsion interaction, where $H_{ij}$ represents the strength of the repulsion and $\eta_{ij}$ is chosen to be 11 for Ge-Ge, 9 for Ge-Se and 7 for Se-Se interactions[1–3]. The $H_{ij}$ factor is given by

$$H_{ij} = A_{ij}(\sigma_i + \sigma_j)^{\eta_{ij}},$$

where $\sigma_i$ is the ionic radius of the of the $i$th ions and $A_{ij}$ is the repulsion strength.

The original charges for Ge and Se were +1.320 and -0.66 respectively, but these values will not keep the total charge of the system to zero, especially for the Ge$_{0.5}$Se$_{0.5}$ and for the Ge$_{0.4}$Se$_{0.6}$ systems. For this reason, the new charges have been computed by the Rappe and Goddard's QEq scheme[5] using the GULP package[6] on the Ge$_{0.5}$Se$_{0.5}$ crystalline system. Charges for the Ge$_{0.4}$Se$_{0.6}$ and for the Ge$_{0.6}$Se$_{0.4}$ systems have been obtained by a linear fit of the Ge$_{0.5}$Se$_{0.5}$ and GeSe2 charges, ensuring that the overall system was neutrally charged, and they are reported in **Table S1**. The three-body term was omitted for simplicity since the two-body terms describe adequately both short- and medium-range features of chalcogenides[7]. The cut-off radius for long-range interactions was taken to be 16Å, accordingly to the original work of Vashishta et al.[1,8]. These parameters are consistent with the description of previous Ge$_x$Se$_{1-x}$ systems, as reported in Results. As a benchmark to have a further validation of our parameters, we compared the radial distribution function of the Ge$_{0.5}$Se$_{0.5}$ at the liquid phase from our simulation with experimental data obtained from Petri et al. on Ge$_{0.5}$Se$_{0.5}$ at 700K[9]. Results reported in **Figure S1** show a good agreement for the Ge-Se and Se-Se interactions, while a discrepancy is observed for Ge-Ge interactions. In this last case, experimental results show two peaks that converged together in one single peak observed in our simulations. The position of the higher experimental peak coincides with the peak obtained from our simulations. However, it should be noted that experimental data are derived as a fit of the Fourier transform of the structure functions and that the Ge-Ge curve is not trivial to reproduce due to several fluctuations.

**Table S1:** Parameters of the Force Field for the three stoichiometries of $Ge_xSe_{1-x}$.

| | $Q_{Ge}$ | $Q_{Se}$ | Density of amorphous phase ($\frac{g}{cm^3}$) |
|---|---|---|---|
| $Ge_{0.5}Se_{0.5}$ | 0.533 | -0.533 | 5.1 |
| $Ge_{0.4}Se_{0.6}$ | 0.900 | -0.600 | 4.36[10] |
| $Ge_{0.6}Se_{0.4}$ | 0.300 | -0.450 | 5.05 |
| Force Field parameters | | | |
| $H_{ij}(Ge-Ge)$ | | | 16.043 eV |
| $H_{ij}(Ge-Se)$ | | | 2103.20 eV |
| $H_{ij}(Se-Se)$ | | | 4091.08 eV |
| $\alpha_{Ge}$ | | | 0 |
| $\alpha_{Se}$ | | | 7.0 Å$^3$ |
| $\eta_{Ge-Ge}$ | | | 11 |
| $\eta_{Ge-Se}$ | | | 9 |
| $\eta_{Se-Se}$ | | | 7 |
| $A_{ij}$ | | | 249.7 meV |
| $\sigma_{Ge}$ | | | 0.73 Å |
| $\sigma_{Se}$ | | | 2.00 Å |
| $r_{4s}$ | | | 4.43 Å |

**S2. Complementary structural analysis of $Ge_xSe_{1-x}$ compounds**

**Figure S1** reports the g(r) of the three systems considered in the main text in the liquid phase (T=1500K) and the comparison with the experimental work of Petri et al.[9]. It can be observed that Ge-Se and Se-Se interactions match well with our simulations, while Ge-Ge the two peaks are gathered together in a single broader peak. This small discrepancy is due to the lower level of accuracy of classical MD simulations with respect to *ab initio* simulations, but the position of the main peak is perfectly reproduced. Notably, there is no general consensus between experimental and theoretical works both on $GeSe_2$ and GeSe about the shape of the g(r) for Ge-Ge[9,11–13].

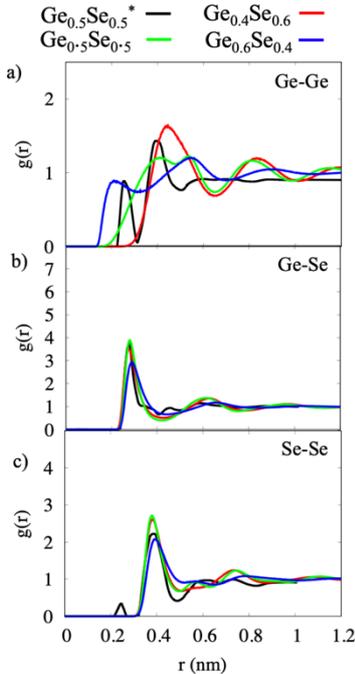

**Figure S1**: panels a) to c) the radial distribution function of the liquid phase at 1500K for each $Ge_xSe_{1-x}$ system. Experimental data for $Ge_{0.5}Se_{0.5}$ (black line) are retrieved from the work of Petri et al.[9] obtained by the fit of the Fourier Transform of the Structure Factor of the liquid phase.

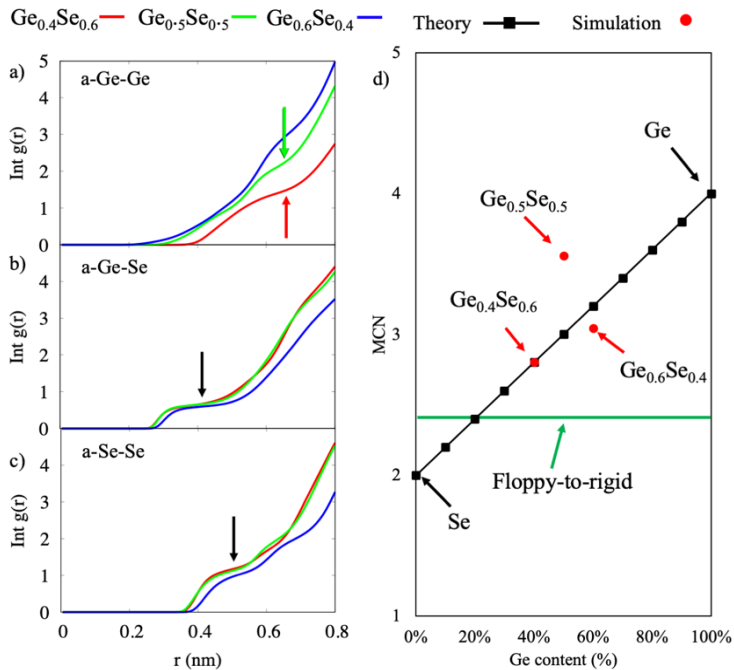

**Figure S2**: Panels a) to c) report the g(r) integral for the amorphous phase showing the coordination numbers. Arrows highlight the plateau of the g(r) integral showing the first neighboring shell. Panel d) the MCN obtained from the empirical formula, black dotted line, and computed from our simulation, red dots.

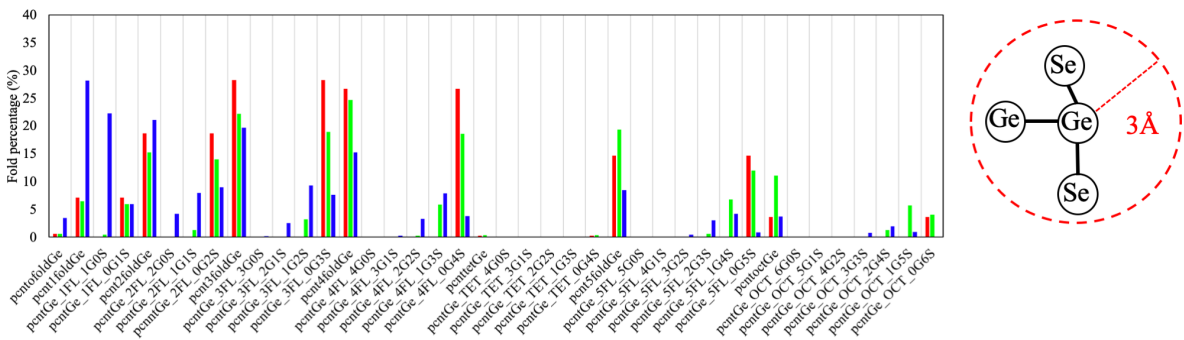

**Figure S3:** fold statistics for central Ge atoms. The label follows this rule: central atom name (pcntGe)_fold(nFL)_number of Ge neighboring atoms(xG)_number of Se neighboring atoms(yS). As an example, the label pcntGe_2FL_0G_2S means that a central Ge atom is 2-folded to 0 Ge and to 2 Se atoms. On the right a sketch of the connections of a central Ge atom with both Ge and Se within a cutoff of 3Å, i.e. pcntGe_3FL_1G2S.

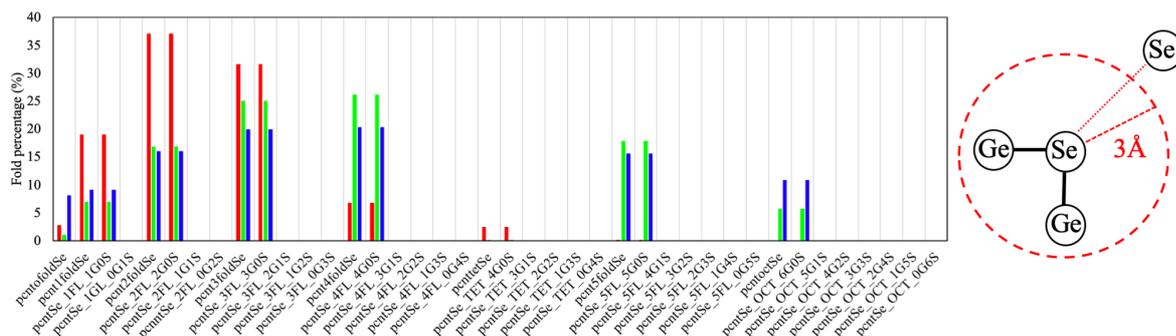

**Figure S4:** fold statistics for central Se atoms. The label follows this rule: central atom name (pcntSe)_fold(nFL)_number of Ge neighboring atoms(xG)_number of Se neighboring atoms(yS). As an example, the label pcntSe_3FL_3G_0S means that a central Se atom is 3-folded to 3 Ge and to 0 Se atoms. On the right a sketch of the connections of a central Se atom with Ge only within a cutoff of 3Å, i.e. pcntSe_2FL_2G0S.

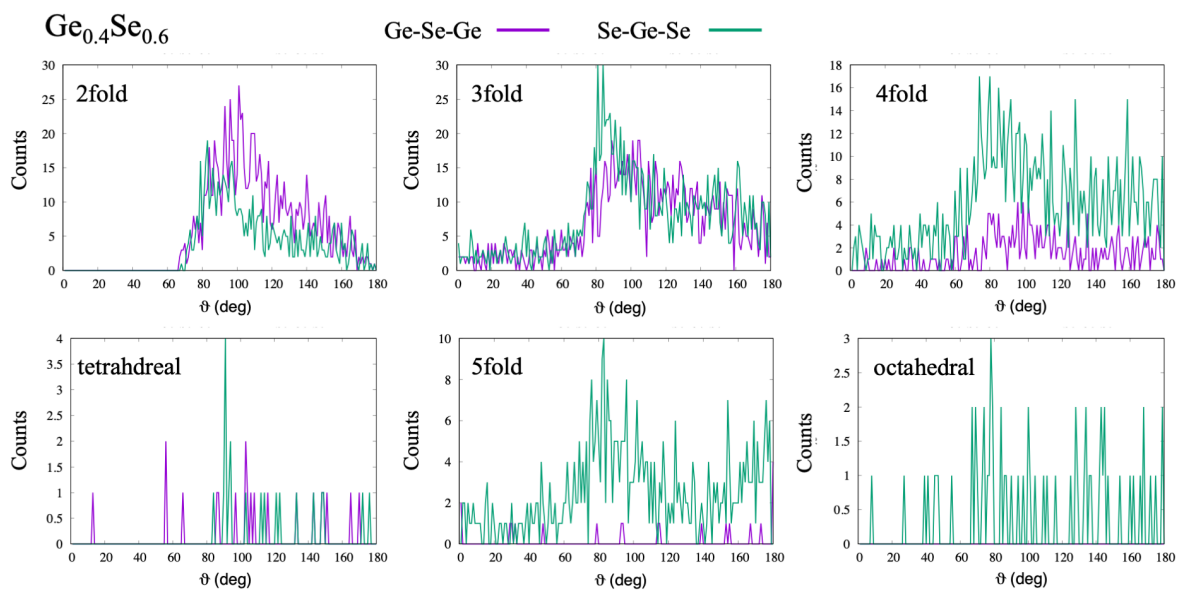

**Figure S5:** decomposition of angle distributions in Ge-Se-Ge and Se-Ge-Se for each fold for the $Ge_{0.4}Se_{0.6}$ system.

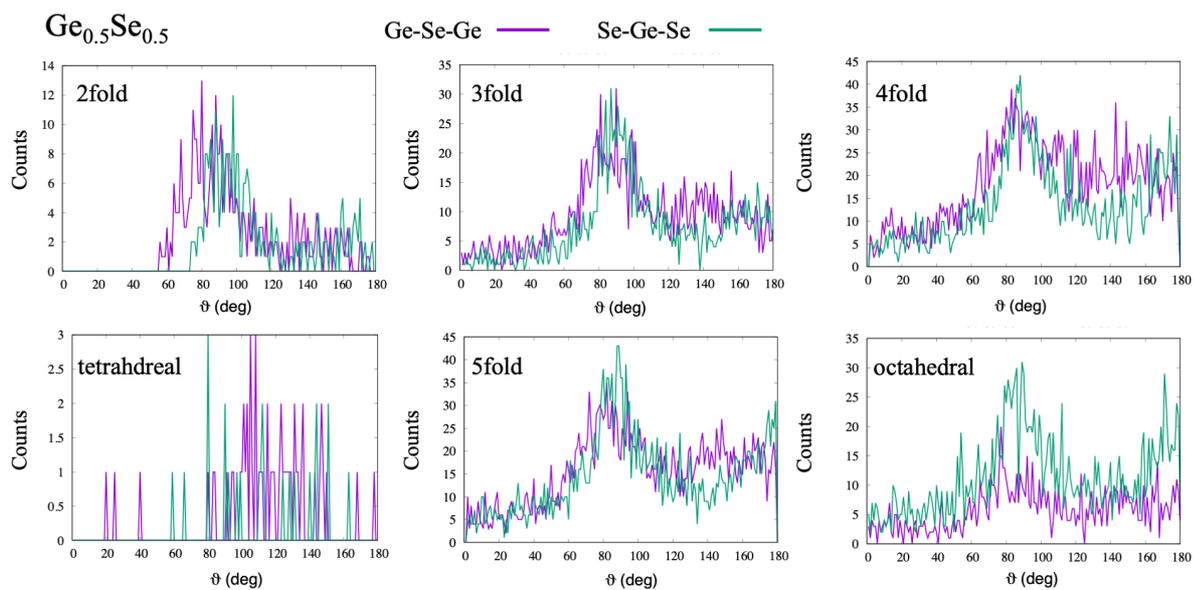

**Figure S6:** decomposition of angle distributions in Ge-Se-Ge and Se-Ge-Se for each fold for the $Ge_{0.5}Se_{0.5}$ system.

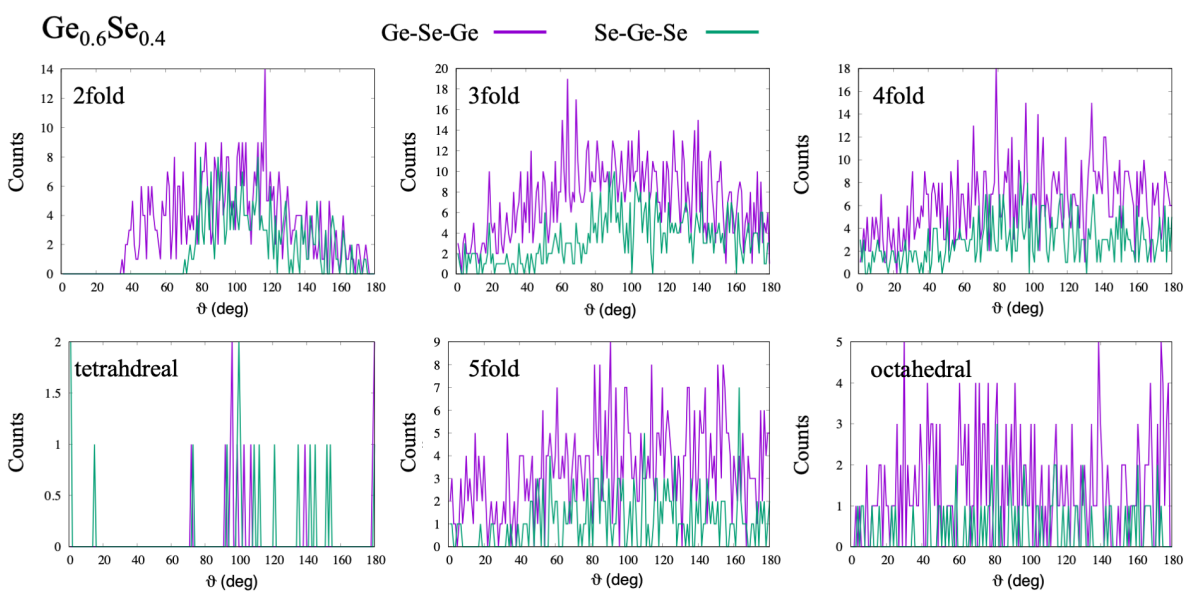

**Figure S7:** decomposition of angle distributions in Ge-Se-Ge and Se-Ge-Se for each fold for the $Ge_{0.5}Se_{0.5}$ system.

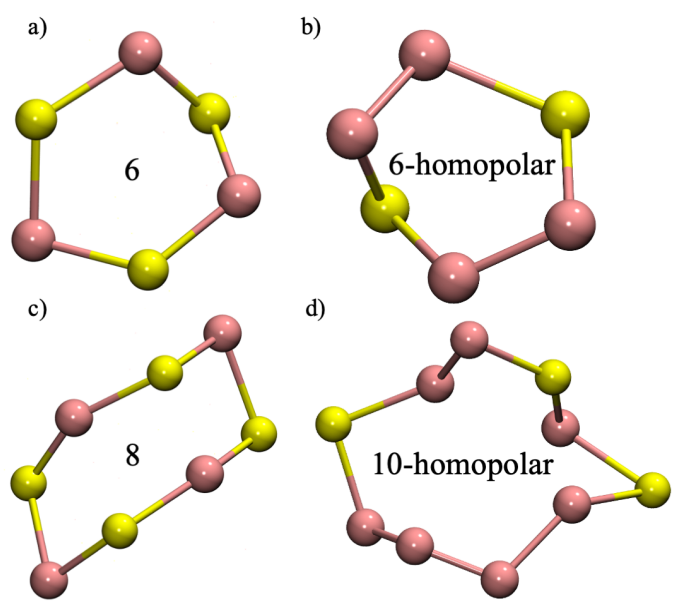

**Figure S8:** some typical structures of rings of different diameter: a 6-member ring in panel a), 6-homopolar in b), 8 in c) and 10-homopolar in d). Se atoms are colored in yellow and Ge atoms in pink.

**S3. Ab initio molecular dynamics simulation**

In order to gain insight on the structural properties of the $Ge_xSe_{1-x}$ compounds, we performed an ab initio molecular dynamics simulation of the $Ge_{0.5}Se_{0.5}$ system. Time evolution is simulated using ab initio Car-Parrinello (CP) molecular dynamics[14], as implemented in the Quantum Espresso package[15]. The electronic structure is described within the Density functional theory (DFT) approach. The exchange-correlation functional is described by using the vdW-DF2-B86R[16] formulation of generalized gradient approximation, which includes vdW-DF corrections to describe non-bonding interactions. Atomic potentials are described by ultrasoft pseudopotential of Vanderbilt type[17]. Single particle wavefunctions (charge) are expanded in plane waves up to a kinetic energy cutoff of 30 Ry (300 Ry). The fictitious mass associated with the electronic orbital degrees of freedom is set to 600 au. The time step for molecular dynamics evolution is dt=0.17 fs. The ionic temperature is controlled by a chain of two Nose'-Hoover thermostats[18,19] with frequencies $\omega_1$ =50 THz and $\omega_2$=25 THz, respectively. To fix the undesired heat transfer from the ionic to the fictitious-electronic degrees of freedom, we coupled a further Nose'-Hoover thermostat ($\omega_{el}$=100 THz) to the electronic orbital degrees of freedom in the Car-Parrinello Lagrangian[20].

Initial configuration is obtained from an amorphous model already published in the litterature[21], which includes 336 atoms (118 Ge and 118 Se). We performed a melting-and-quenching amorphization cycle, by heating the system to 1500K for 15 ps, followed by a cell optimization (Parrinello-Rahman) at 1200K for 10 ps and by a cooling step to 300K, with a rate of 100K/ps. After quenching, a production run has been conducted at 300K for 15ps.

The g(r) plots, shown in **Figure S9**, reproduce the characteristics features of the $Ge_{0.5}Se_{0.5}$ system, presented in the main text.

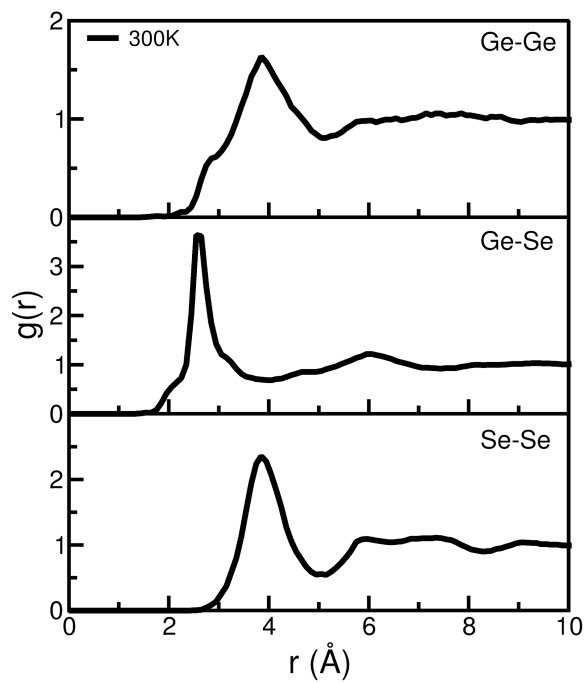

**Figure S9**: radial distribution function, g(r), for the amorphous phase at 300K for each $Ge_{0.5}Se_{0.5}$ system, from ab initio simulation.

The structural analysis of the resulting ab initio CP model and the comparison with the MD one are summarized in **Table S2** and in **Figure S10**. Ab initio CP model has a predominant distribution of low-folded structures (0-3 fold) which are characteristic of the melt phase, while higher folded aggregation are minor components, in evident contrast with the classical MD results at the same temperature. Computational limitations on system size and simulation time (especially quenching step), do not allow the ab initio model to proper cool down and aggregate in highly-folded structures, typical of the glass phase.

|  | 0-FOLD | 1-FOLD | 2-FOLD | 3-FOLD | 4-FOLD | TETRA-HEDRAL | 5-FOLD | OCTA-HEDRAL |
|---|---|---|---|---|---|---|---|---|
| **CP(300K)** | 0,77% | 21,93% | 39,10% | 33,85% | 2,84% | 1,22% | 0,29% | 0,00% |
| **MD (300K)** | 0.9% | 6.7% | 16.0% | 23.6% | 25.6% | 0.3% | 18.5% | 8.4% |

**Table S2**: Average percentage of folded structure from ab initio CP and classical MD simulations at 300K.

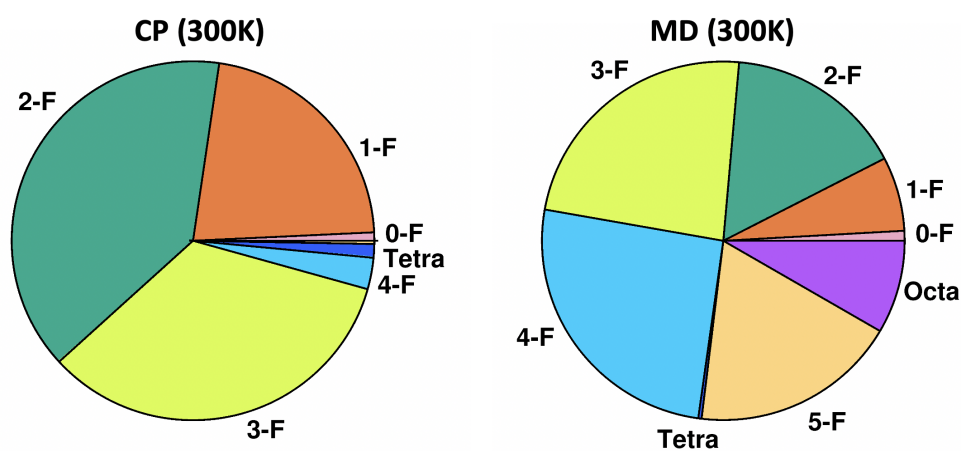

**Figure S10**: Pie chart distribution of folded structures from ab initio CP and classical MD simulations at 300K.